\documentclass{ws-ijmpe}

\def\lsim{\raise0.3ex\hbox{$<$\kern-0.75em\raise-1.1ex\hbox{$\sim$}}}
\def\gsim{\raise0.3ex\hbox{$>$\kern-0.75em\raise-1.1ex\hbox{$\sim$}}}

\def\mean#1{\left<#1\right>}
\def\Journal#1#2#3#4{{#1}{\bf#2} (#4) #3}

\def\NPB{{\it Nucl. Phys.}~{\bf B}}

\def\PRL{\it Phys. Rev. Lett.\ }
\def\PRD{{\it Phys. Rev.}~{\bf D}}
\def\PRC{{\it Phys. Rev.}~{\bf C}}

\def\ARNPS{{\it Ann. Rev. Nucl. Part. Sci.\ }}

\begin{document}

\markboth{M. J. Tannenbaum}{Why the $x_E$ distribution...}

\catchline{}{}{}{}{}

\title{WHY THE $x_E$ DISTRIBUTION  TRIGGERED BY A LEADING PARTICLE DOES NOT MEASURE THE FRAGMENTATION FUNCTION BUT DOES MEASURE THE RATIO OF THE TRANSVERSE MOMENTA OF THE AWAY-SIDE JET TO THE TRIGGER-SIDE JET. \footnote{Research supported by U. S. Department of Energy, DE-AC02-98CH10886.}
}

\author{\footnotesize M. J. TANNENBAUM}

\address{Physics Department, 510c,\\
Brookhaven National Laboratory,\\
Upton, NY 11973-5000, USA\\
mjt@bnl.gov}\maketitle

\begin{history}
\received{(received date)}
\revised{(revised date)}
\end{history}

\begin{abstract}
  Hard-scattering of point-like constituents (or partons) in p-p collisions was discovered at the CERN-ISR~\cite{Darriulat} in 1972 by measurements utilizing inclusive single or pairs of hadrons with large transverse momentum ($p_T$). Due to the steeply falling power-law ${p}_T$  spectrum of the hard-scattered partons, the inclusive single particle (e.g. $\pi^0$) $p_{T_t}$ spectrum from parton fragmentation to a jet is dominated by trigger fragments with large $\mean{z_t}\sim 0.7-0.8$, where $z_t=p_{T_t}/p_{T{\rm jet}}$ is the fragmentation variable. It was generally assumed, following Feynman, Field and Fox~\cite{FFF}, as shown by data from the CERN-ISR experiments, that the $p_{T_a}$ distribution of away side hadrons from a single particle trigger [with $p_{T_t}$], corrected for $\mean{z_t}$, would be the same as that from a jet-trigger and follow the same fragmentation function as observed in $e^+ e^-$  or DIS. PHENIX~\cite{ppg029} attempted to measure the fragmentation function from the away side $x_E\sim p_{T_a}/p_{T_t}$ distribution of charged particles triggered by a $\pi^0$ in p-p collisions at RHIC and showed by explicit calculation that the $x_E$ distribution is actually quite insensitive to the fragmentation function.  Illustrations of the original arguments and ISR results will be presented. Then the lack of sensitivity to the fragmentation function will be  explained, and an analytic formula for the $x_E$ distribution given, in terms of incomplete Gamma functions, for the case where the fragmentation function is exponential. The away-side distribution in this formulation has the nice property that it both exhibits $x_E$ scaling and is directly sensitive to the ratio of the away jet $\hat{p}_{T_a}$ to that of the trigger jet, $\hat{p}_{T_t}$, and thus can be used, for example, to measure the relative energy loss of the two jets from a hard-scattering which escape from the medium in A+A collisions. Comparisons of the analytical formula to RHIC measurements will be presented, including data from STAR~\cite{STARPRL95FQW,STARPRL97}  and PHENIX~\cite{ppg029,ppg039}, leading to some interesting conclusions.

\end{abstract}
\section{Hard Scattering in p-p collisions at the CERN-ISR} 
Following the discovery of hard-scattering at the CERN-ISR~\cite{Darriulat} by the observation of an unexpectedly large yield of particles with large transverse momentum $(p_T)$, which proved that the quarks of DIS were strongly interacting, the attention of experimenters turned to measuring the predicted di-jet structure of the hard-scattering events using two-particle correlations. 
\begin{figure}[!htb]\vspace*{-0.25cm}
\begin{center}
\begin{tabular}{cc}
\psfig{file=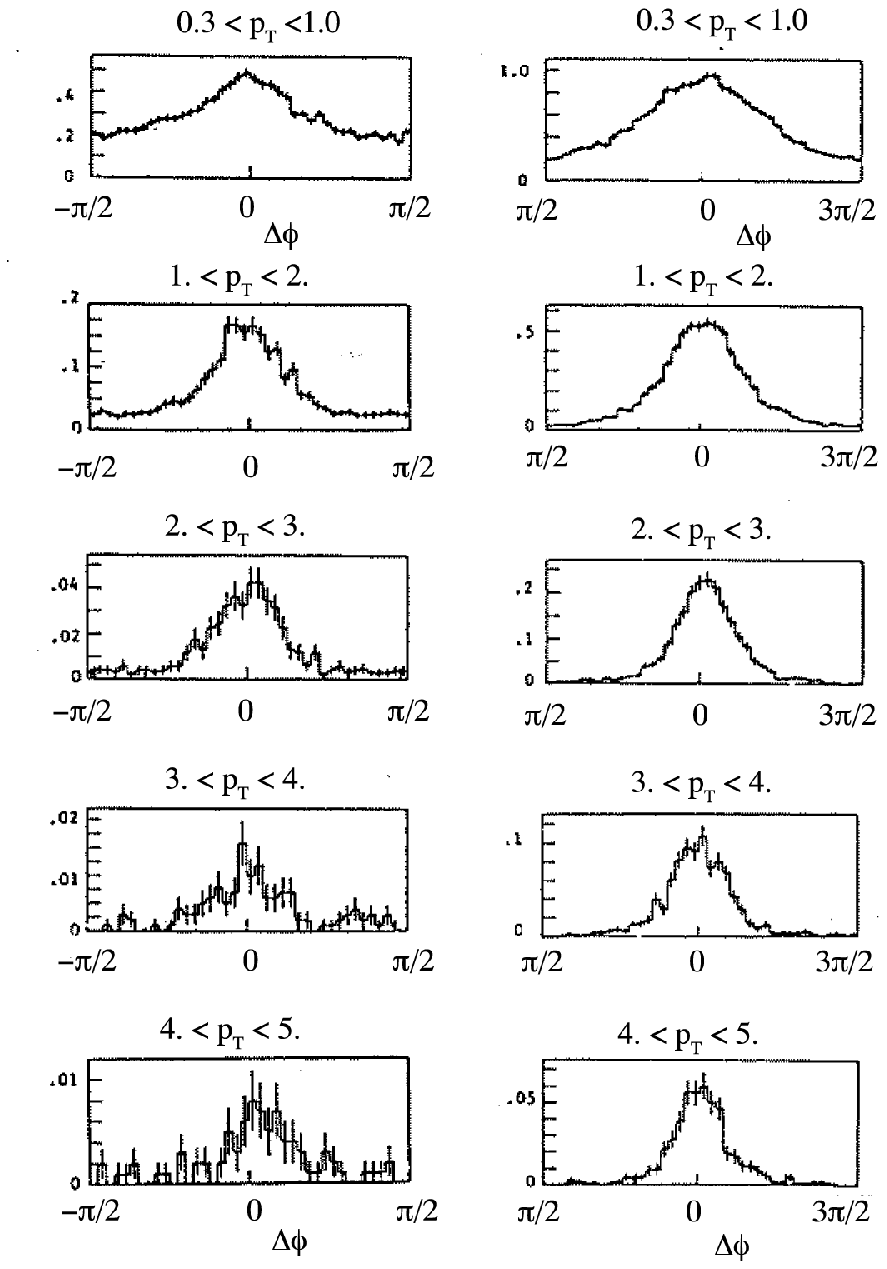,width=0.4\linewidth,height=0.5\linewidth}&\hspace*{-0.044\linewidth}
\psfig{file=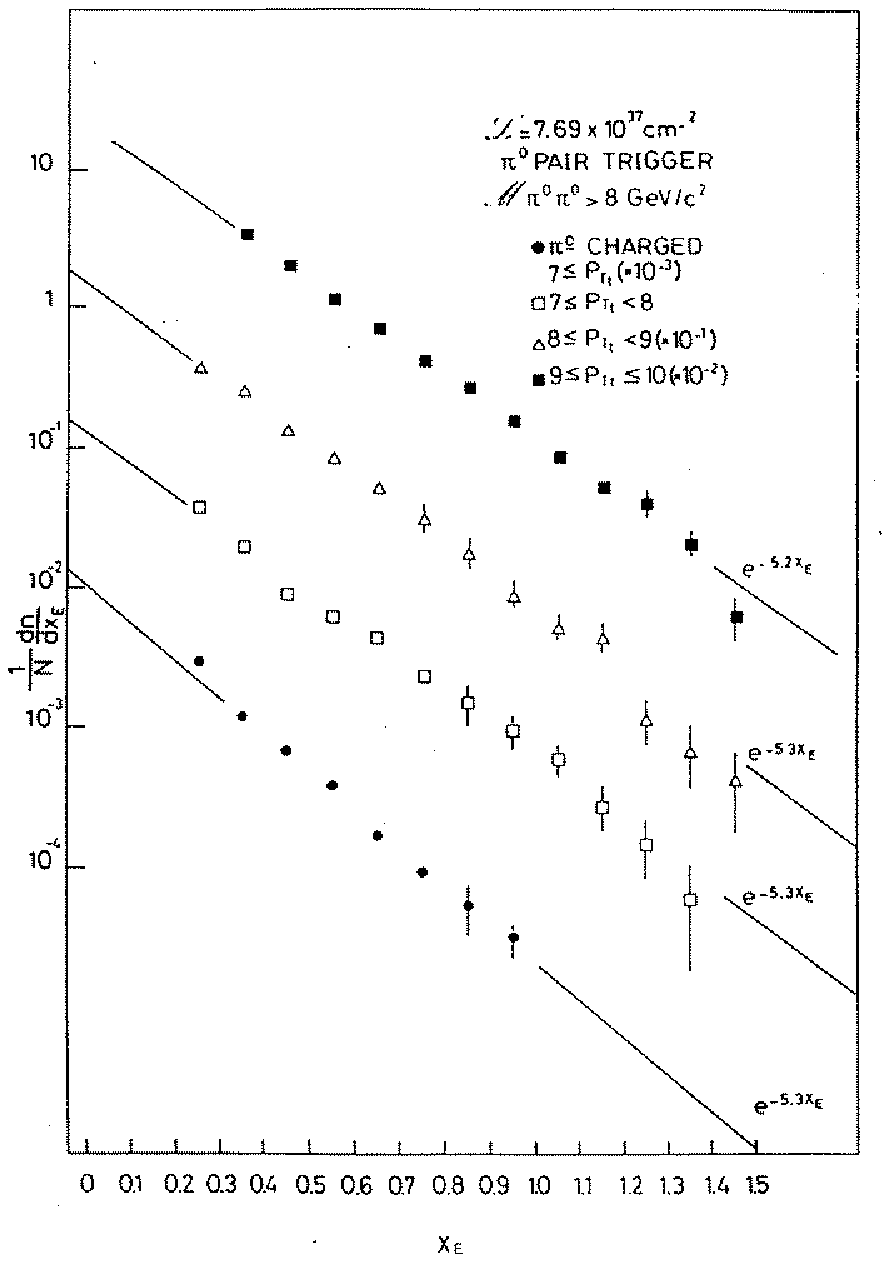,width=0.4\linewidth,height=0.5\linewidth}
\end{tabular}
\end{center}\vspace*{-0.15in}
\caption[]{a) (left) trigger side b) (center) away-side correlations of charged particles with indicated $p_T$ for $\pi^0$ triggers with $p_{T_t} > 7$ GeV/c  . b) (right) $x_E$ distributions from this data .    \label{fig:mjt-ccorazi}}
\end{figure}\vspace*{-0.5cm}
    The CCOR experiment~\cite{Angelis79}, using a $\pi^0$ trigger with transverse momentum $p_{T_t} > 7$ GeV/c,  was the first to provide associated charged particle measurement with full and uniform acceptance over the entire azimuth, with pseudorapidity coverage $-0.7\leq\eta\leq +0.7$, so that the jet structure of high $p_T$ scattering could be easily seen and measured (Fig.~\ref{fig:mjt-ccorazi}a,b). In all cases strong correlation peaks on flat backgrounds are clearly visible, indicating di-jet structure.  The small variation of the widths of the peaks opposite to the trigger for $p_{T}>1$ GeV/c (Fig.~\ref{fig:mjt-ccorazi}b) indicates out-of-plane activity from the individual fragments of jets.

\begin{figure}[!htb]
\begin{center}\vspace*{-0.12in}
\begin{tabular}{cc}
\psfig{file=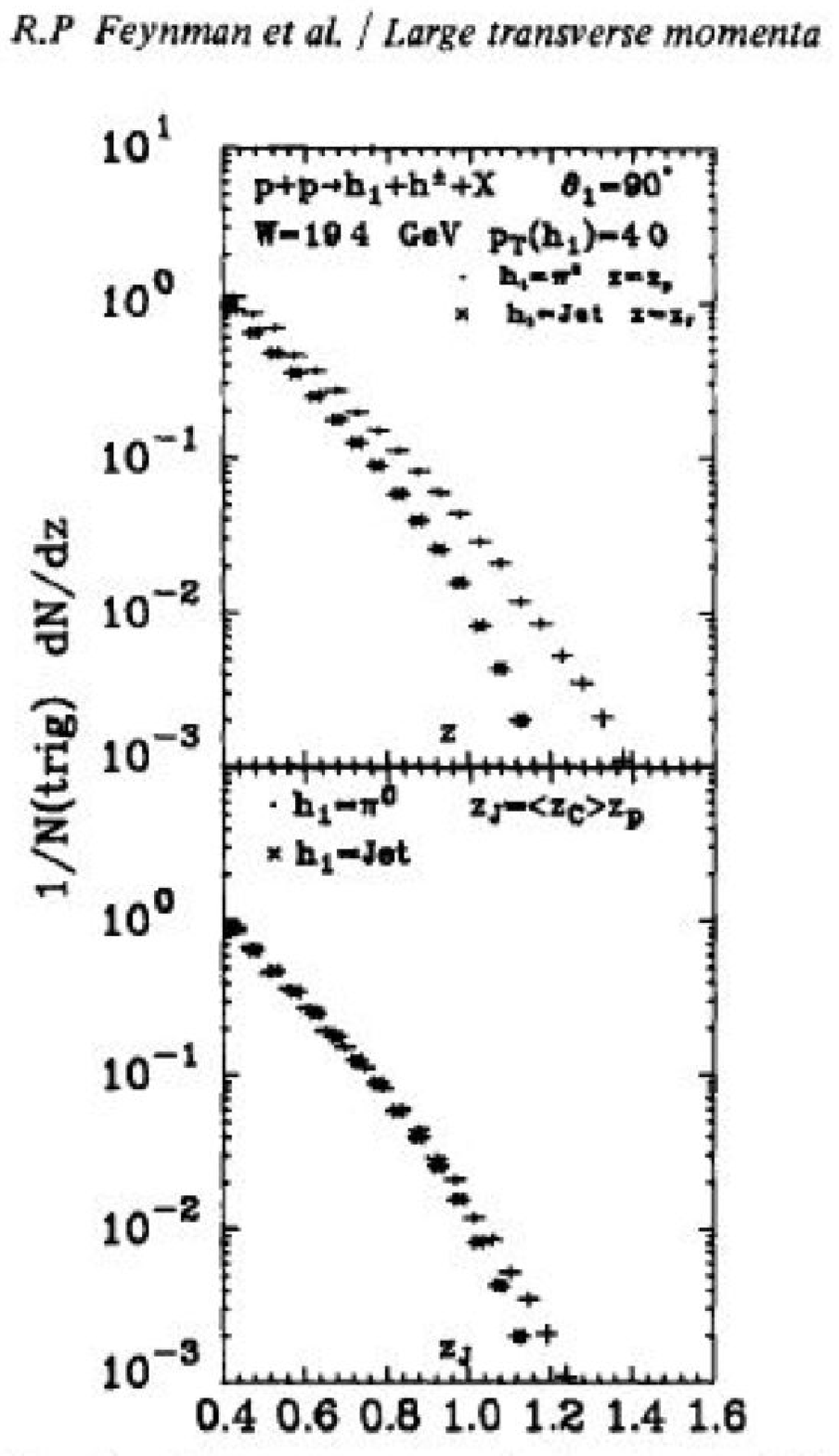,width=0.47\linewidth,height=0.50\linewidth}&\hspace*{-0.10\linewidth}
\psfig{file=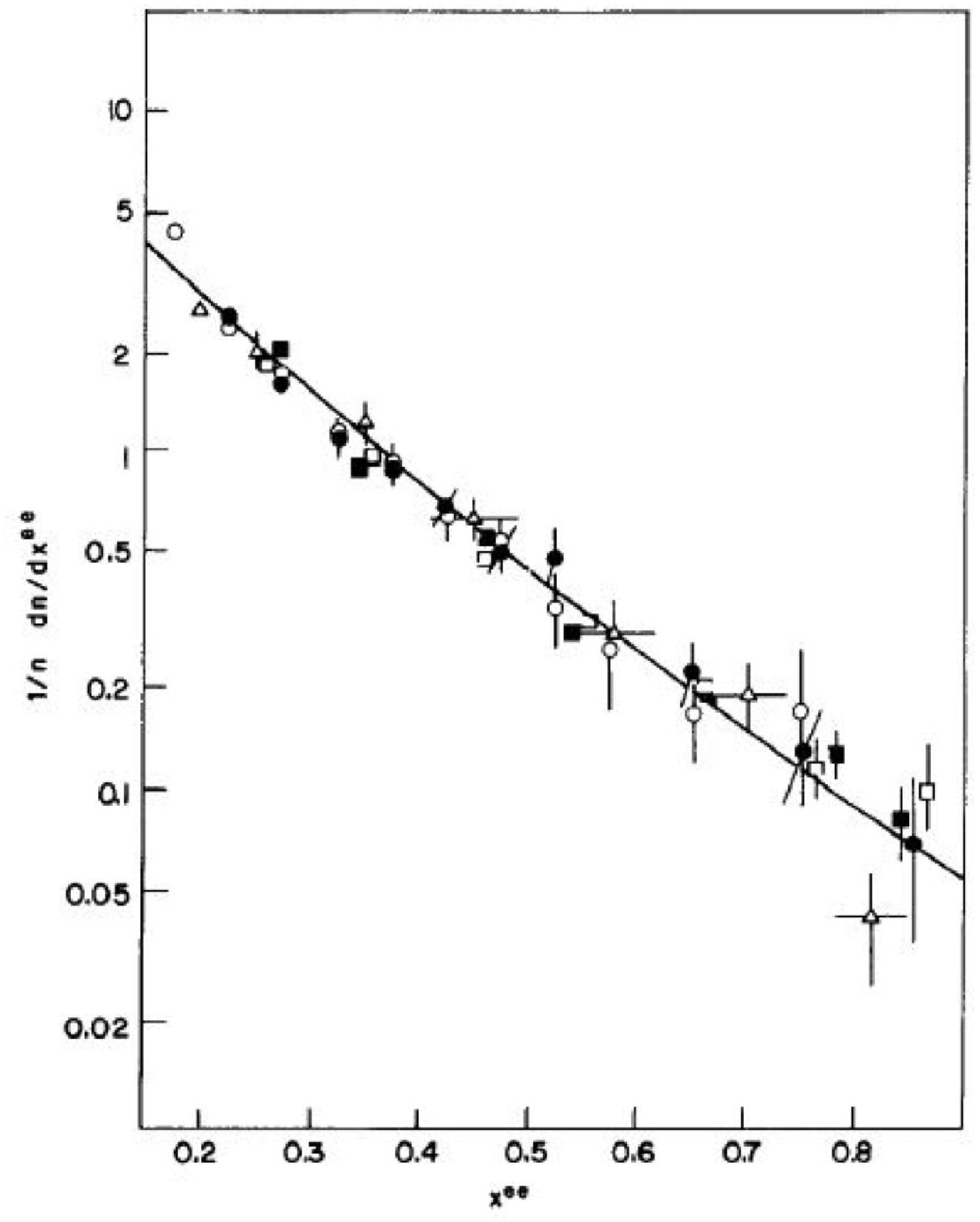,width=0.39\linewidth}
\end{tabular}
\end{center}\vspace*{-0.15in}
\caption[]{a) (left) [top] Comparison~\cite{FFF} of away side charged hadron distribution triggered by a $\pi^0$ or a jet, where $z_{\pi^0}=x_E$ and $z_j=p_{T_a}/p_{T{\rm jet}}$. [bottom] same distributions with $\pi^0$ plotted vs $z'_j=\mean{z_t} x_E$. b) (right) Jet fragmentation functions~\cite{Darriulat} from $\nu$-p, $e^+ e^-$ and p-p collisions (Fig.~\ref{fig:mjt-ccorazi}c).    \label{fig:xxx2}}\vspace*{-1.5cm}
\end{figure}

      	Following the methods of previous CERN-ISR experiments~\cite{Darriulat,FFF}, the away jet azimuthal angular distributions  of Fig.~\ref{fig:mjt-ccorazi}b, which were thought to be unbiased, were analyzed in terms of the two variables: $p_{\rm out}=p_T \sin(\Delta\phi)$, the out-of-plane transverse momentum of a track;  
 and $x_E$, where:\\ 
\begin{minipage}[b]{0.49\linewidth}
\begin{equation*}	
x_E=\frac{-\vec{p}_T\cdot \vec{p}_{T_t}}{|p_{T_t}|^2}=\frac{-p_T \cos(\Delta\phi)}{p_{T_t}}\simeq \frac {z}{z_{t}}  
\nonumber
\end{equation*}
\vspace*{0.06in}
\end{minipage}
\begin{minipage}[b]{0.49\linewidth} 
\vspace*{0.06in}
\psfig{file=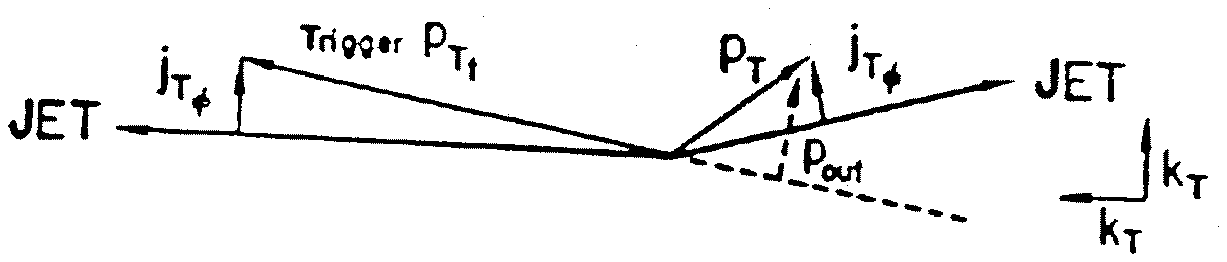, width=\linewidth}
\vspace*{-0.12in}
\label{fig:mjt-poutxe}
\end{minipage}\vspace*{-0.12in}
$z_{t}\simeq p_{T_t}/p_{T{\rm jet}}$ is the fragmentation variable of the trigger jet, and $z$ is the fragmentation variable of the away jet. Note that $x_E$ would equal the fragmenation fraction $z$ of the away jet, for $z_{t}\rightarrow 1$, if the trigger and away jets balanced transverse momentum. 
The $x_E$ distributions~\cite{Angelis79} for the data of Fig.~\ref{fig:mjt-ccorazi}b are shown in Fig.~\ref{fig:mjt-ccorazi}c and show the fragmentation behavior expected at the time, $e^{-6z}\sim e^{-6 x_E \langle z_{t}\rangle}$. It was generally assumed, following the seminal article of Feynman, Field and Fox~\cite{FFF}, that the $p_{T_a}$ distribution of away side hadrons from a single particle trigger [with $p_{T_t}$], corrected for $\mean{z_t}$, would be the same as that from a jet-trigger (Fig.~\ref{fig:xxx2}a) and follow the same fragmentation function as observed in $e^+ e^-$  or DIS, as apparently shown by ISR measurements~\cite{Darriulat} (Fig.~\ref{fig:xxx2}b). 

\section{Hard scattering in p-p and A+A at RHIC}
\begin{figure}[!htb]
\begin{center}\vspace*{-0.22in}
\begin{tabular}{cc}
\psfig{file=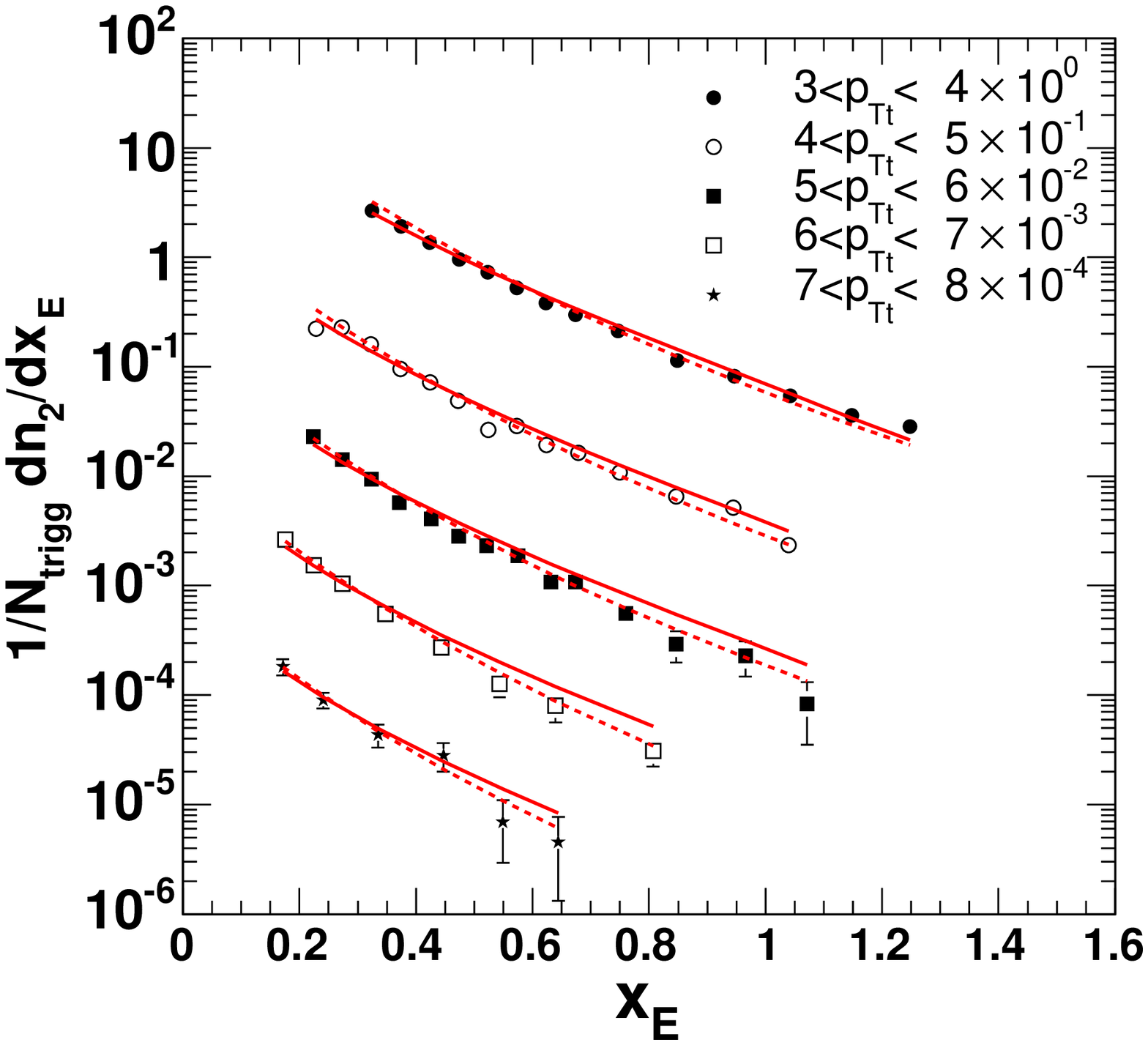,width=0.4\linewidth}&\hspace*{-0.044\linewidth}
\psfig{file=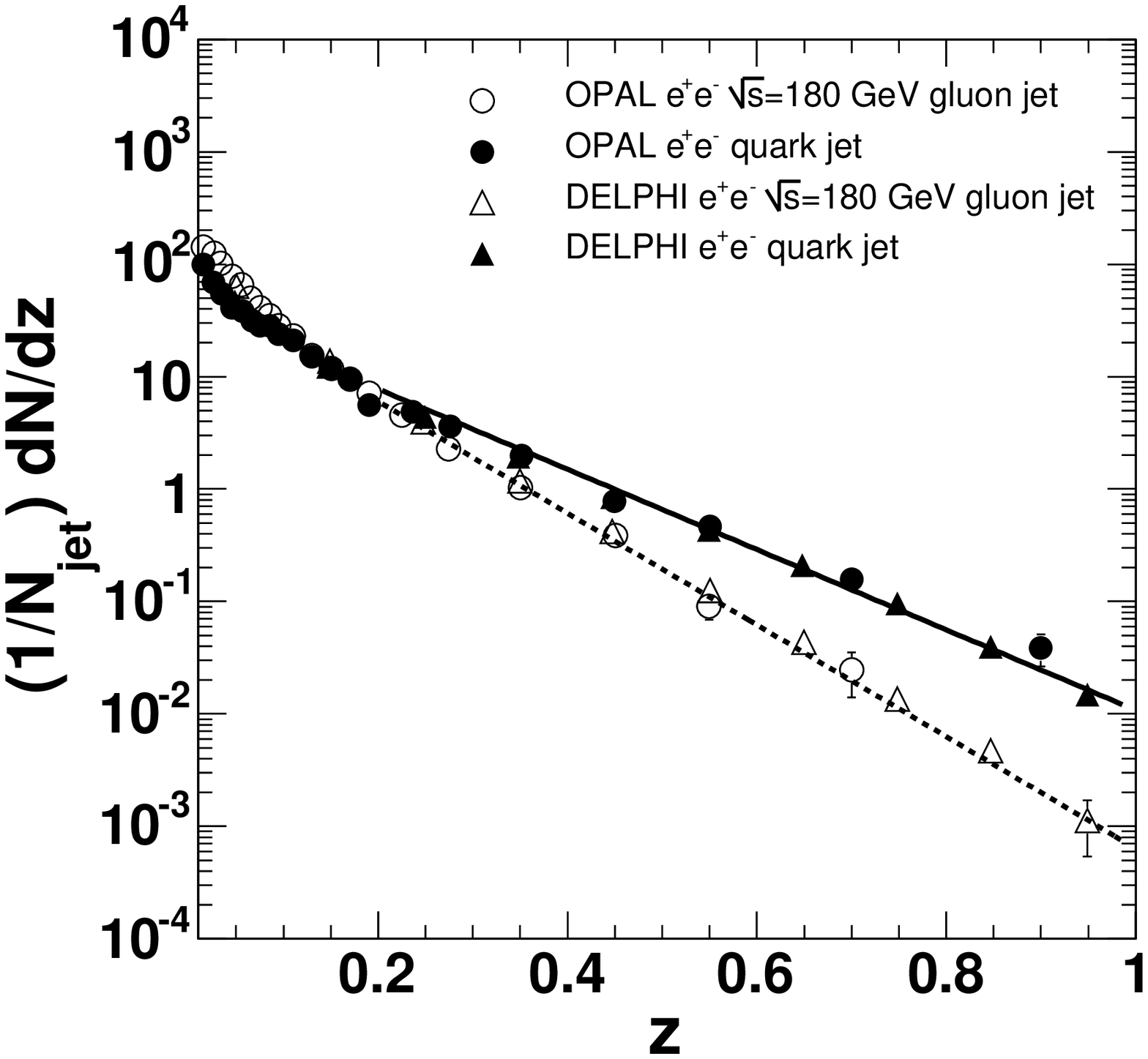,width=0.4\linewidth}
\end{tabular}
\end{center}\vspace*{-0.25in}
\caption[]{a) (left) $x_E$ distributions~\cite{ppg029} together with calculations using fragmentation functions from LEP; b) (right) LEP fragmentation functions. See Ref.~\refcite{ppg029} for details .    \label{fig:xxx3}}\vspace*{-0.12in}
\end{figure}

    PHENIX~\cite{ppg029} attempted to measure the mean net transverse momentum of the di-jet ($\mean{p_{T_{\rm pair}}}=\sqrt{2}\mean{k_T}$) in p-p collisions at RHIC, where $k_T$ (see above) represents the out-of-plane activity of the hard-scattering~\cite{FFF,Darriulat}. This requires the knowledge of $\mean{z_t}$ of the trigger $\pi^0$, which PHENIX attempted to calculate using a fragmentation function derived from the measured $x_E$ distributions (Fig.~\ref{fig:xxx3}a).  
It didn't work. Finally, it was found that starting with either the quark $\approx \exp (-8.2 \cdot z)$ or the gluon $\approx \exp (-11.4 \cdot z)$ fragmentation functions from LEP (Fig.~\ref{fig:xxx3}b solid and dotted lines), which are quite different in shape, the results obtained for the $x_E$ distributions (solid and dotted lines on Fig.~\ref{fig:xxx3}a) do not differ significantly! Although nobody had noticed this for nearly 30 years, the reason turned out to be quite simple. 

	The joint probability for a fragment pion with $p_{T_t}=z_t\hat{p}_{T_t}$, originating from a parton with $\hat{p}_{T_t}$, and a fragment pion with $p_{T_a}=z_a\hat{p}_{T_a}$, originating from the other parton in the hard-scattered pair with $\hat{p}_{T_a}$, where $\hat{p}_{T_a}/\hat{p}_{T_t}\equiv \hat{x}_h\lsim\ 1$ due to $k_T$ smearing~\cite{FFF,ppg029}  is:  
\begin{equation}
 { {d^3\sigma_{\pi} (\hat{p}_{T_t},z_t,z_a) }\over {\hat{p}_{T_t}d\hat{p}_{T_t}dz_t dz_a}}=
 {{d\sigma_q}\over {\hat{p}_{T_t}d\hat{p}_{T_t}}}\times D^{\pi}_q(z_t) \times D^{\pi}_q(z_a) \qquad , 
\label{eq:mjt-zaztgivenq}
\end{equation}
where ${{d\sigma_{q} }/{\hat{p}_{T_t} d\hat{p}_{T_t}}}$ is the $k_T$-smeared final-state scattered-parton invariant spectrum and $D^{\pi}_q (z_t)$  
represents the fragmentation function.   
 Changing variables from $\hat{p}_{T_t}$, $z_t$ to $p_{T_t}$, $z_t$, and from $z_a$ to $p_{T_a}$ gives:
\begin{equation} 
{ {d^3\sigma_{\pi} }\over {d{p}_{T_t}dz_t dp_{T_a}}}= {1\over {\hat{x}_{\rm h}\, p_{T_t}}}
 {{d\sigma_q}\over {d({p}_{T_t}/z_t})}\,D^{\pi}_q(z_t) D^{\pi}_q({{z_t p_{T_a}} \over {\hat{x}_{\rm h} p_{T_t}}}) \qquad .
 \label{eq:mikes_dsig}
 \end{equation}
The $x_E$ distribution can be found by integrating over $dz_t$ in Eq.~\ref{eq:mikes_dsig} and dividing by the inclusive $p_{T_t}$ cross section.   
With no assumptions other than a power law for the parton $\hat{p}_{T_t}$ distribution (${{d\sigma_{q} }/{\hat{p}_{T_t} d\hat{p}_{T_t}}}= A \hat{p}_{T_t}^{-n}$) , an exponential fragmentation function ($D^{\pi}_q (z)=B e^{-bz}$) and constant $\hat{x}_h$, the result for the $x_E$ distribution in the collinear limit, where, $p_{T_a}=x_E p_{T_t}$ is: 
   \begin{equation}
\left.{dP_{\pi} \over dx_E}\right|_{p_{T_t}}= {1\over\hat{x}_h} {B\over b} {1\over
{(1+ {x_E \over{\hat{x}_h}})^{n}}}  
\, { {\left[ \Gamma({n},b' x_{T_t}) - \Gamma (n, b'{\hat{x}_h\over x_E}) \right] } \over { \left[ \Gamma({n-1},b x_{T_t}) - \Gamma (n-1, b) \right]}} 
\qquad , 
\label{eq:ans1_condxe}
\end{equation}
where $b'=b(1+ {p_{T_a} / {\hat{x}_h p_{T_t}}}) $ and $\Gamma(a,x)$ is the upper incomplete Gamma function, where $\Gamma(a,0)=\Gamma(a)$ and $\Gamma(a)=(a-1)!$ for $a$ an integer. 

A reasonable approximation is obtained by taking the upper limit of the integral over $z_t$ to infinity and the lower limit to zero, so that the ratio of incomplete Gamma functions in Eq.~\ref{eq:ans1_condxe} becomes equal to $\Gamma(n)/\Gamma(n-1)=n-1$ and the $x_E$ distribution takes on a very simple and very interesting form:
	     \begin{equation}
\left.{dP_{\pi} \over dx_E}\right|_{p_{T_t}}\approx {\mean{m}(n-1)}{1\over\hat{x}_h} {1\over
{(1+ {x_E \over{\hat{x}_h}})^{n}}} \, \qquad . 
\label{eq:condxe2}
\end{equation}

\begin{figure}[!htb]
\begin{center}
\begin{tabular}{cc}
\psfig{file=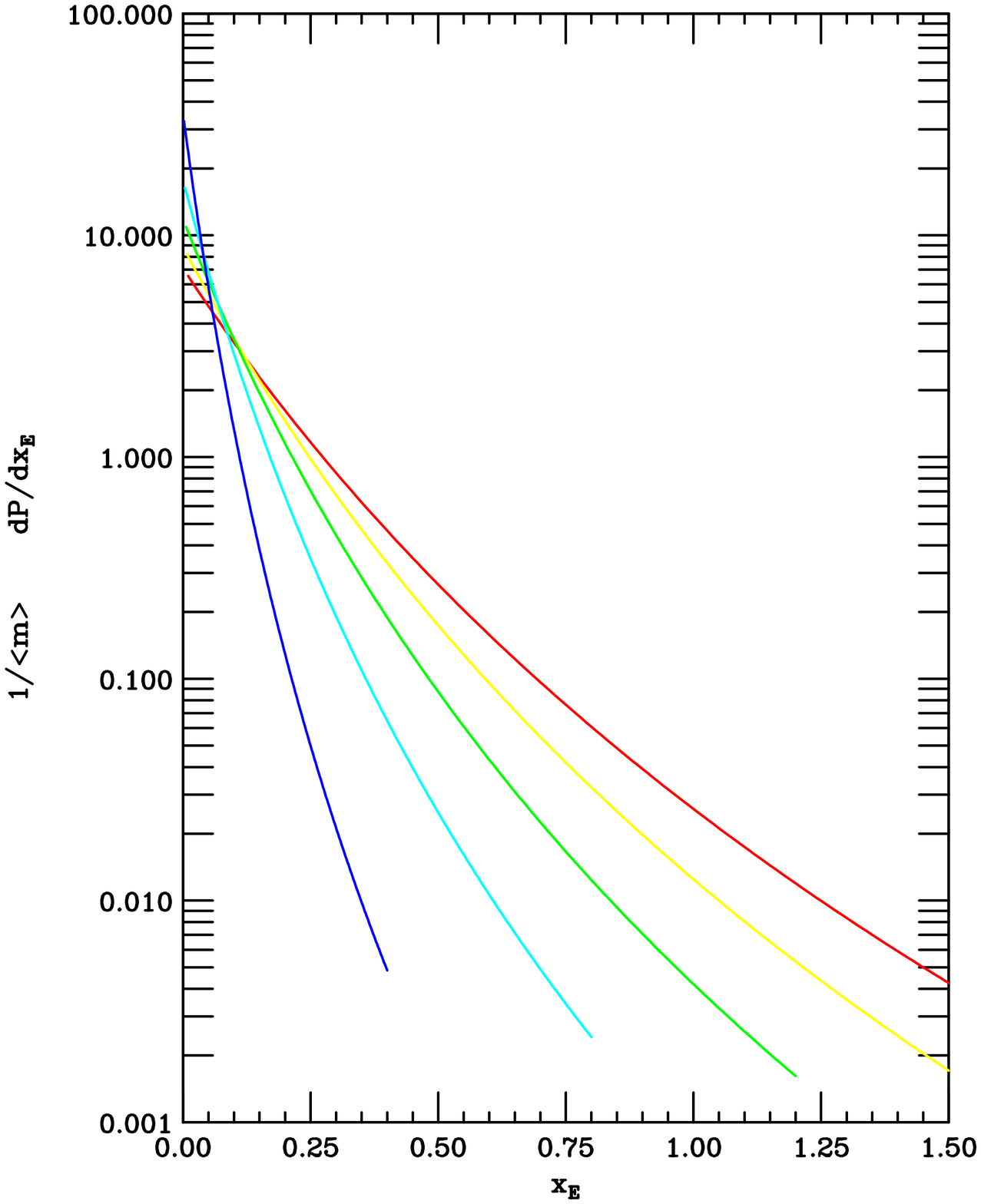,width=0.45\linewidth,height=0.49\linewidth}&\hspace*{-0.04\linewidth}
\psfig{file=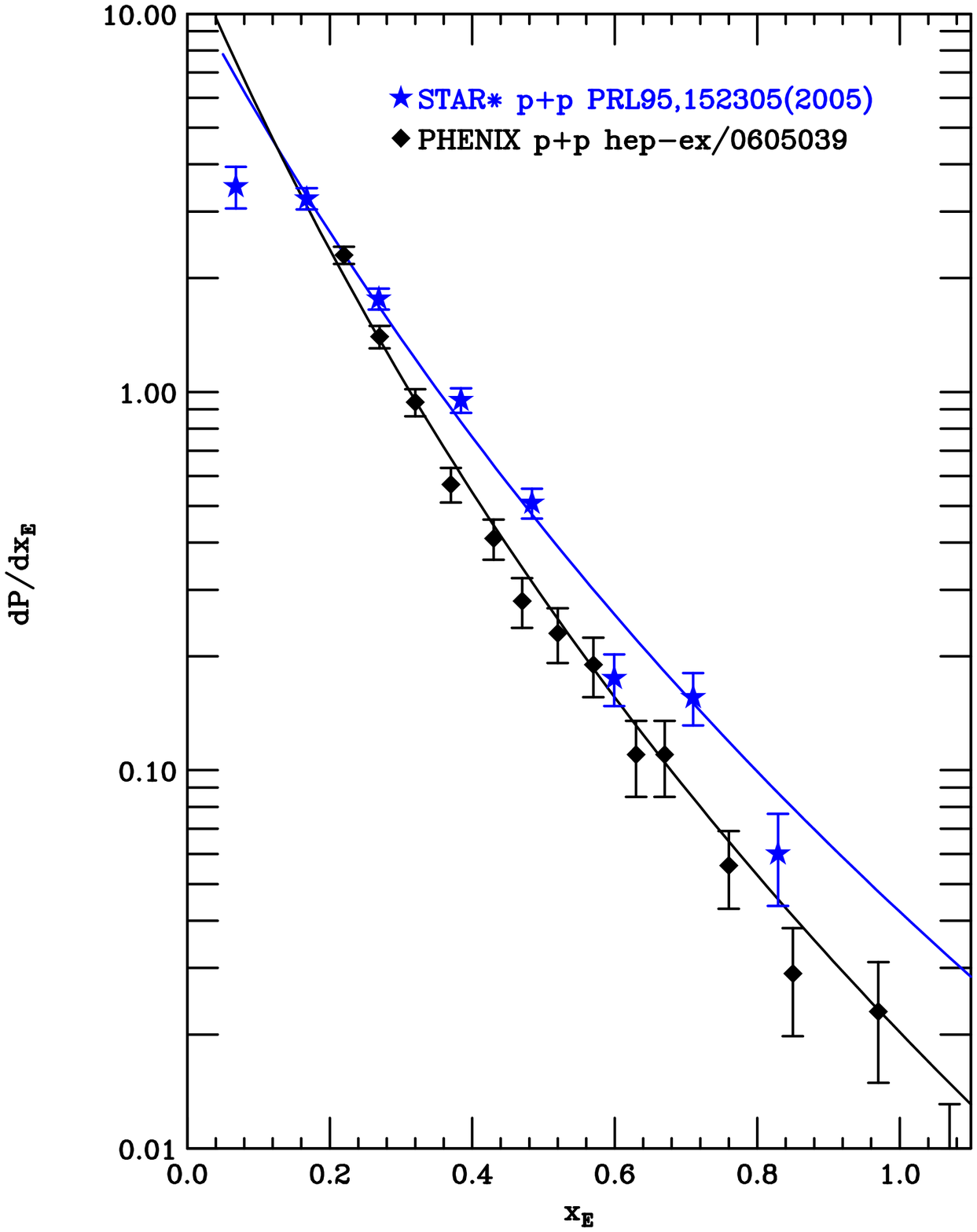,width=0.45\linewidth,height=0.49\linewidth}
\end{tabular}
\end{center}\vspace*{-0.15in}
\caption[]{a) (left) Eq.~\ref{eq:condxe2} divided by $\langle m\rangle$. Curves are for $\hat{x}_h=$1.0 (red), 0.8, 0.6, 0.4, 0.2 (blue). b) (right) Eq.~\ref{eq:condxe2} with $\hat{x}_h$ adjusted to match PHENIX~\cite{ppg029} and STAR~\cite{STARPRL95FQW} p-p data.    \label{fig:xxx4}}
\end{figure}

Equation~\ref{eq:condxe2} has very nice properties:
i) the dominant term is the Hagedorn function $1/(1+x_E/\hat{x}_h)^n$ so that Eq.~\ref{eq:condxe2} exhibits $x_E$-scaling in the variable $x_E/\hat{x}_h$ (see Fig.~\ref{fig:xxx4}a); ii) the shape of the $x_E$ distribution is given by the power $n$ of the partonic and inclusive single particle transverse momentum spectra and does not depend on the exponential slope of the fragmentation function; iii) the only dependence on the fragmentation function is that the integral of the $x_E$ distribution (from zero to infinity) is equal to 
$\mean{m}$, the mean multiplicity of the unbiased away-jet. 

	The reason that the $x_E$ distribution is not very sensitive to the fragmentation function is that the integral over $z_t$ for fixed $p_{T_t}$ and $p_{T_a}$ (Eq.~\ref{eq:mikes_dsig}) is actually an integral over the jet transverse momentum $\hat{p}_{T_t}$. However since both the trigger and away jets are always roughly equal and opposite in transverse momentum, integrating over $\hat{p}_{T_t}$ simultaneously integrates over $\hat{p}_{T_a}$, and thus also integrates over the away jet fragmentation function. This can be seen directly by the presence of $z_t$ in both the same and away fragmentation functions in Eq.~\ref{eq:mikes_dsig}, so that the integral over $z_t$ integrates over both fragmentation functions simultaneously. 
\section{Application to A+A Collisions}
Equation~\ref{eq:condxe2} is simple but powerful. 
 It relates the ratio of the transverse momenta of the away and trigger particles, $p_{T_a}/p_{T_t}\approx x_E$, which is measured, to the ratio of the transverse momenta of the away to the trigger jet, $\hat{p}_{T_a}/\hat{p}_{T_t}$, which can thus be deduced. Although derived for p-p collisions, Eq.~\ref{eq:condxe2} should work just as well in A+A collisions since the only assumptions are independent fragmentation of the trigger and away-jets with the same exponential fragmentation function and a power-law parton $\hat{p}_{T_t}$ distribution. The only other (and weakest) assumption is that $\hat{x}_h$ is constant for fixed $p_{T_t}$ as a function of $x_E$. Thus in A+A collisions, Eq.~\ref{eq:condxe2} for the $x_E$ distribution provides a method of measuring the ratio $\hat{x}_h=\hat{p}_{T_a}/\hat{p}_{T_t}$ and hence the relative energy loss of the away to the same side jet assuming that both jets fragment outside the medium with the same fragmentation function as in p-p collisions. 
 
\begin{figure}[!htb]
\begin{center}
\begin{tabular}{cc}
\psfig{file=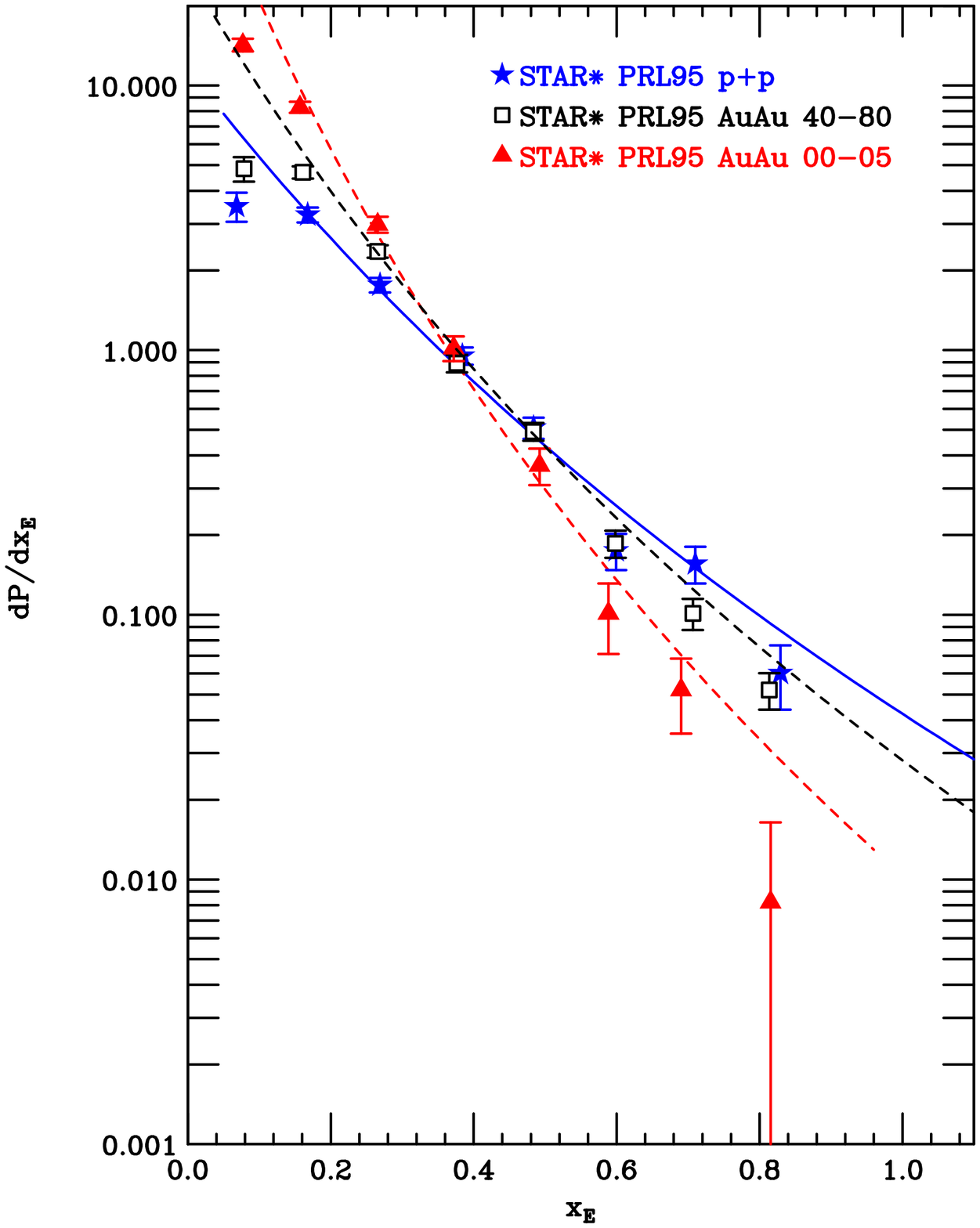,width=0.45\linewidth,height=0.49\linewidth}&\hspace*{-0.03\linewidth}
\psfig{file=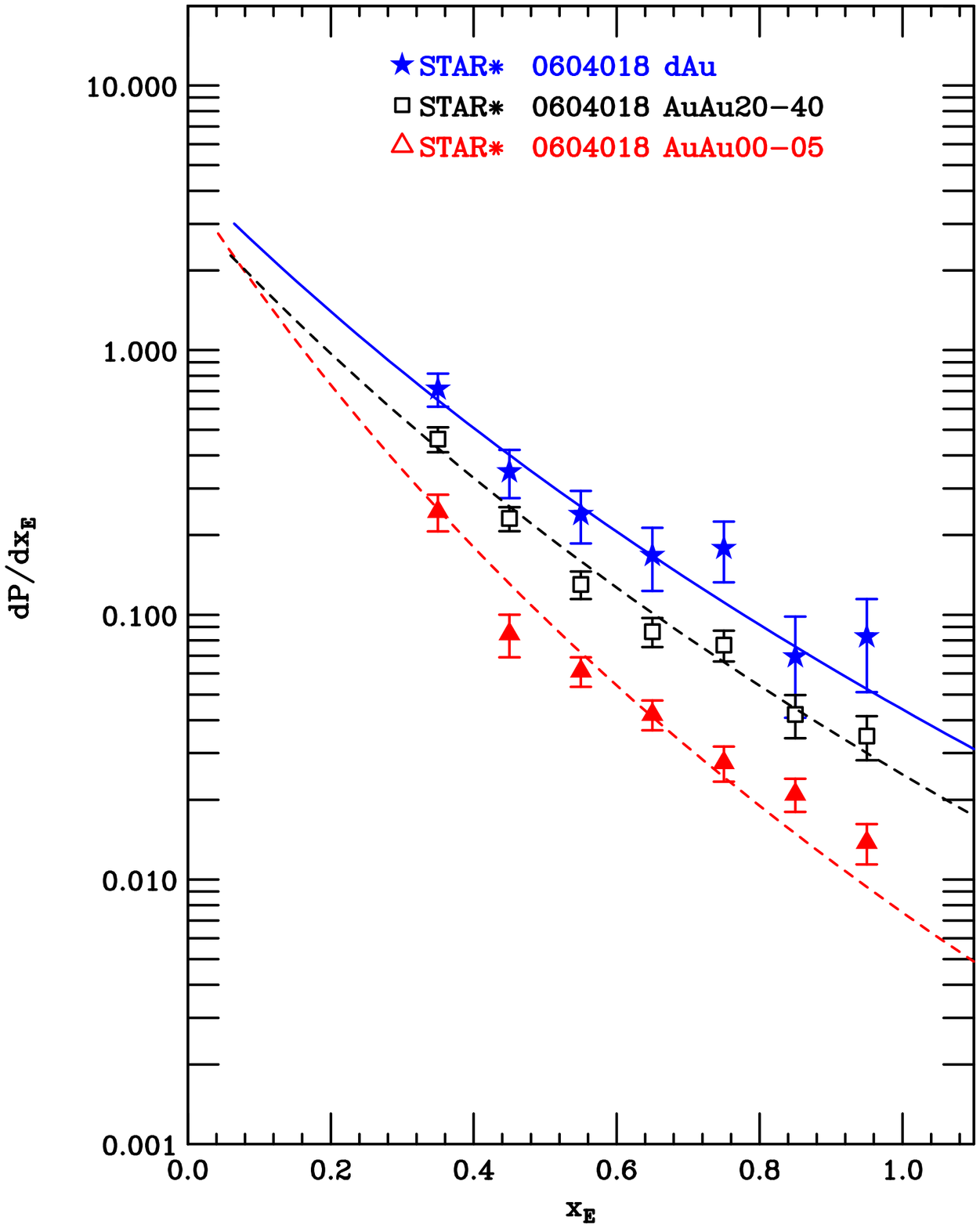,width=0.45\linewidth,height=0.49\linewidth}
\end{tabular}
\end{center}\vspace*{-0.15in}
\caption[]{Eq.~\ref{eq:condxe2} adjusted to match STAR data: a) (left) Ref.~\refcite{STARPRL95FQW} . b) (right) Ref.~\refcite{STARPRL97} .    \label{fig:xxx5}}
\end{figure}

      To test the validity of Eq.~\ref{eq:condxe2} we first apply it to the PHENIX~\cite{ppg029} and STAR~\cite{STARPRL95FQW} p-p data, with the fixed value of $n=8.1$ measured for $\pi^0$ in p-p~\cite{ppg054} by adjusting the normalization and the scale $\hat{x}_h$ to agree with the data. The values are $\hat{x}_h=0.8$ for PHENIX~\cite{ppg029} in agreement with the calculated value from $k_T$ smearing, and $\hat{x}_h=1$ for STAR~\cite{STARPRL95FQW}. We then follow the same procedure for the STAR Au+Au data~\cite{STARPRL95FQW,STARPRL97} (Fig~\ref{fig:xxx5}). The best values are $\hat{x}_h=1$ for p-p, $\hat{x}_h=0.75$ for Au+Au peripheral, and $\hat{x}_h=0.48$ for Au+Au central (Fig.~\ref{fig:xxx5}a)~\cite{STARPRL95FQW}; and $\hat{x}_h=1.30$ for d+Au, $\hat{x}_h=1.20$ for Au+Au peripheral (20-40\%) and $\hat{x}_h=0.85$ for Au+Au central (0-5\%) (Fig.~\ref{fig:xxx5}b)~\cite{STARPRL97}. Thus, to within the error of the simple scaling analysis, both measurements indicate a clear and steady decrease of the energy of the away jet relative to the trigger jet with increasing centrality, reaching a ratio of $\sim 0.5$ for central Au+Au collisions compared to the p-p or d+Au baseline. Due to the surface bias of the trigger jets, these results indicate that the energy loss of the away jet increases with distance travelled in the medium. It will be very interesting to compare this simple analysis and conclusion to the results and conclusions of full theoretical analyses of the data. Of course the experimental results still must be improved since both the absolute values and the $x_E$ slopes of the two STAR measurements disagree with each other and the slopes for $x_E >0.4$ of all the data in Fig.~\ref{fig:xxx5}b~\cite{STARPRL97} are much flatter than other measurements in p-p and d+Au collisions in the same $p_{T_t}$ range~\cite{ppg039} as reflected in the anomalous  
value of $\hat{x}_h\approx 1.30$ in this range for all systems. 
\end{document}